\documentclass[sigconf,screen]{acmart}

\usepackage{subfigure}
\usepackage{subcaption}
\usepackage{acronym}
\usepackage[table]{xcolor}
\usepackage{booktabs}
\usepackage{multirow}

\title{The Effect of Multi-Lingual and Keyword Adversarial Injection on LLM Relevance Judgment}

\begin{document}

\author{Nguyen Khoi Vo}
% \email{s3891987@rmit.edu.vn}
\affiliation{%
  \institution{RMIT University}
  \city{Melbourne}
  \state{VIC}
  \country{Australia}
}

\author{Tuong Duy Duong}
% \email{s3479994@student.rmit.edu.au}
\affiliation{%
  \institution{RMIT University}
  \city{Melbourne}
  \state{VIC}
  \country{Australia}
}

\author{Mark Sanderson}
% \email{mark.sanderson@rmit.edu.au}
\affiliation{%
  \institution{RMIT University}
  \city{Melbourne}
  \state{VIC}
  \country{Australia}
}

\author{Oleg Zendel}
% \email{oleg.zendel@rmit.edu.au}
\affiliation{%
  \institution{RMIT University}
  \city{Melbourne}
  \state{VIC}
  \country{Australia}
}

\renewcommand{\shortauthors}{Vo et al.}

\begin{abstract}
  Large language models (LLMs) are increasingly being used as automated judges
  for relevance evaluation in information retrieval, yet their robustness to
  adversarial manipulation remains insufficiently understood, particularly in
  multilingual settings. In this work, we investigate the impact of
  cross-lingual prompt injection attacks on LLM-based relevance judgments using
  TREC Deep Learning collections and two open-weight models under established
  prompting frameworks. We examine both instruction-based and content-based
  injection strategies in 8 languages spanning different resource levels. Our
  results demonstrate that multilingual query-based injections are highly
  effective in inflating relevance scores while simultaneously evading existing
  prompt-injection defenses. We further found that, although existing defense
  mechanisms can be modified to mitigate such attacks, these injections can be
  easily adapted to bypass them. These findings highlight a critical gap in
  current defense approaches and demonstrate that language generalization can
  act as an attack vector, underscoring the need for more robust and proactive
  evaluation frameworks for LLM-as-a-judge systems.
\end{abstract}


\copyrightyear{2026}
\acmYear{2026}
\setcopyright{cc}
\setcctype{by}
% remove DOI and ISBN from ACM header
\acmDOI{} \acmISBN{} \acmConference[VulGen'26]{The International Workshop on
  Vulnerabilities in Generative Systems for Information Retrieval}{July 20--24,
  2026}{Melbourne, VIC, Australia} \acmBooktitle{}

\maketitle

\section{Introduction}

Recent advances in LLMs have positioned them as potential alternatives to
traditional human-based relevance judgments, leading to their increasing
adoption as automated judges in information retrieval (IR)
tasks~\cite{dietzPrinciplesGuidelinesUse2025, thomasLargeLanguageModels2024}. As
LLM judges are deployed in evaluation pipelines—including TREC-style benchmarks
and commercial search quality assessment; their susceptibility to adversarial
manipulation carries potential consequences: inflated relevance scores can
distort evaluation outcomes, misguide retrieval system development, and
undermine the integrity of large-scale automated annotation.

% Prior work on adversarial LLM evaluation is disjointed:
% \citet{alaofiLLMsCanBe2025} explore keyword injection in judgments, while
% \citet{cuconasuPowerNoiseRedefining2024} focus on distracting content
% degrading RAG performance. The robustness of these content-based attacks—such
% as query variants or semantic distractions—remains poorly understood,
% particularly regarding cross-lingual transferability
% \cite{thomasSystemComparisonUsing2025a}.

Prior work on adversarial manipulation of LLM-based evaluation has been limited.
Most studies focus on instruction-based prompt injection (e.g., ``ignore
previous instructions'')~\cite{li2025survey}. In contrast, only a small number
of works have examined \emph{content-based} manipulation. In particular,
\citet{alaofiLLMsCanBe2025} shows that inserting query keywords into passages can
fool LLM-based relevance judgments, while
\citet{cuconasuPowerNoiseRedefining2024} demonstrates that introducing
distracting content can significantly degrade retrieval-augmented generation
(RAG) performance. However, these two lines of work remain largely disconnected:
the first focuses on \emph{keyword-based injection}, while the second focuses on
\emph{performance degradation} in RAG systems, without evaluating adversarial
implications for LLM-as-a-judge settings. Furthermore, limited attention has
been given to richer forms of content manipulation, such as query variants or
semantically similar but irrelevant text, and to their transferability across
languages. In particular, content-based injection strategies -- such as the
inclusion of query keywords, phrases, or their variants -- resemble keyword
stuffing and black-hat SEO practices, yet their robustness and generality remain
insufficiently understood. This gap is particularly concerning given recent
findings by \citet{thomasSystemComparisonUsing2025a}, which suggest that
LLM-based evaluation can make relevance judgments across languages.
This raises the question of whether content-based manipulations can also
transfer across languages and remain effective under multilingual settings.

In this work, we conduct a preliminary study on multilingual content-based
injections, including query keywords and variants, in LLM-based relevance
judgment settings. We also evaluate existing defense mechanisms and find that
they fail to mitigate these attacks. Overall, our findings identify
content-based multilingual injection as an underexplored attack surface for
LLM-as-a-judge systems and highlight the need for more robust evaluation and
defense methods.

% A comprehensive study by \citet{thomasLargeLanguageModels2024} found that
% translating prompts and passages has minimal impact on LLM-based relevance
% judgments, suggesting strong cross-lingual generalisability. However, this
% also raises concerns for adversarial robustness, as injections written in one
% language may remain effective when translated into another, enabling scalable
% multilingual attacks. The study is limited to a single proprietary model
% (GPT-4) and four relatively high-resource languages, leaving the effects on
% smaller models and lower-resource languages unclear. Evidence from
% low-resource settings is mixed: \citet{jesus_exploring_2024} shows LLaMA3 can
% perform relevance judgments in Tetun despite limited data, while
% \citet{nazi_evaluation_2025} reports inconsistencies and increased
% hallucination in GPT-4o on Bengali tasks. A recent review by
% \citet{correia2026systematicliteraturereviewllm} further identifies
% multilingual prompting as an emerging attack surface, highlighting “mismatched
% generalisation”, where safety behaviours learned primarily from high-resource
% languages fail to reliably transfer to others.
\begin{table*}[tb]
  \centering
  \captionof{table}{False-positive (FP) and false-negative (FN) rates for TREC-DL 2022 after multilingual Query Phrase (QP) and Instruction (Instruct) injections. "BASE" denotes the uninjected baseline. \vspace*{-0.1cm}}
  \label{tab:fp_fn_total_vulgen}
  \resizebox{0.98\textwidth}{!}{%
    \begin{tabular}{llcccccccccccccccccc}
      \toprule
      \multirow{2}{*}{\textbf{Model}} &          & \multicolumn{2}{c}{\textbf{BASE}}             & \multicolumn{2}{c}{\textbf{AR}}             & \multicolumn{2}{c}{\textbf{ENG}}             & \multicolumn{2}{c}{\textbf{GA}}             & \multicolumn{2}{c}{\textbf{HE}}             & \multicolumn{2}{c}{\textbf{RU}}             & \multicolumn{2}{c}{\textbf{SW}}             & \multicolumn{2}{c}{\textbf{TH}}             & \multicolumn{2}{c}{\textbf{VI}}             \\
      \cmidrule(lr){3-4}\cmidrule(lr){5-6}\cmidrule(lr){7-8}\cmidrule(lr){9-10}\cmidrule(lr){11-12}\cmidrule(lr){13-14}\cmidrule(lr){15-16}\cmidrule(lr){17-18}\cmidrule(lr){19-20}
                                      &          & FP                                          & FN                                         & FP                                          & FN                                         & FP                                          & FN                                         & FP                                          & FN                                         & FP                                          & FN                                         & FP                                          & FN                                         & FP                                          & FN                                         & FP                                          & FN                                         & FP                                          & FN                                         \\
      \midrule

      \multirow{2}{*}{\textbf{GPT-OSS}} & QP       & \cellcolor[HTML]{BEE5B8} \color{black} 22\% & \cellcolor[HTML]{ABD0E6} \color{black} 3\% & \cellcolor[HTML]{37A055} \color{white} 28\% & \cellcolor[HTML]{ABD0E6} \color{black} 3\% & \cellcolor[HTML]{46AE60} \color{black} 27\% & \cellcolor[HTML]{6AAED6} \color{black} 4\% & \cellcolor[HTML]{73C476} \color{black} 25\% & \cellcolor[HTML]{6AAED6} \color{black} 4\% & \cellcolor[HTML]{46AE60} \color{black} 27\% & \cellcolor[HTML]{6AAED6} \color{black} 4\% & \cellcolor[HTML]{238B45} \color{white} 30\% & \cellcolor[HTML]{ABD0E6} \color{black} 3\% & \cellcolor[HTML]{5DB96B} \color{black} 26\% & \cellcolor[HTML]{6AAED6} \color{black} 4\% & \cellcolor[HTML]{238B45} \color{white} 30\% & \cellcolor[HTML]{ABD0E6} \color{black} 3\% & \cellcolor[HTML]{0B7734} \color{white} 31\% & \cellcolor[HTML]{ABD0E6} \color{black} 3\% \\
                                       & Instruct & \cellcolor[HTML]{FFFFFF} \color{black} -- & \cellcolor[HTML]{FFFFFF} \color{black} -- & \cellcolor[HTML]{73C476} \color{black} 25\% & \cellcolor[HTML]{ABD0E6} \color{black} 3\% & \cellcolor[HTML]{00441B} \color{white} 69\% & \cellcolor[HTML]{F7FBFF} \color{black} 0\% & \cellcolor[HTML]{DBF1D6} \color{black} 21\% & \cellcolor[HTML]{6AAED6} \color{black} 4\% & \cellcolor[HTML]{37A055} \color{white} 28\% & \cellcolor[HTML]{D6E6F4} \color{black} 2\% & \cellcolor[HTML]{00441B} \color{white} 38\% & \cellcolor[HTML]{F7FBFF} \color{black} 1\% & \cellcolor[HTML]{EFF9EC} \color{black} 18\% & \cellcolor[HTML]{105BA4} \color{white} 6\% & \cellcolor[HTML]{BEE5B8} \color{black} 22\% & \cellcolor[HTML]{ABD0E6} \color{black} 3\% & \cellcolor[HTML]{37A055} \color{white} 28\% & \cellcolor[HTML]{D6E6F4} \color{black} 2\% \\

      \midrule

      \multirow{2}{*}{\textbf{QWEN}} & QP       & \cellcolor[HTML]{5DB96B} \color{black} 26\% & \cellcolor[HTML]{ABD0E6} \color{black} 3\% & \cellcolor[HTML]{005A24} \color{white} 35\% & \cellcolor[HTML]{D6E6F4} \color{black} 2\% & \cellcolor[HTML]{00441B} \color{white} 38\% & \cellcolor[HTML]{F7FBFF} \color{black} 1\% & \cellcolor[HTML]{0B7734} \color{white} 31\% & \cellcolor[HTML]{ABD0E6} \color{black} 3\% & \cellcolor[HTML]{00682A} \color{white} 34\% & \cellcolor[HTML]{D6E6F4} \color{black} 2\% & \cellcolor[HTML]{005A24} \color{white} 35\% & \cellcolor[HTML]{D6E6F4} \color{black} 2\% & \cellcolor[HTML]{238B45} \color{white} 30\% & \cellcolor[HTML]{ABD0E6} \color{black} 3\% & \cellcolor[HTML]{005A24} \color{white} 35\% & \cellcolor[HTML]{D6E6F4} \color{black} 2\% & \cellcolor[HTML]{005A24} \color{white} 36\% & \cellcolor[HTML]{D6E6F4} \color{black} 2\% \\
                                       & Instruct & \cellcolor[HTML]{FFFFFF} \color{black} -- & \cellcolor[HTML]{FFFFFF} \color{black} -- & \cellcolor[HTML]{00441B} \color{white} 50\% & \cellcolor[HTML]{F7FBFF} \color{black} 0\% & \cellcolor[HTML]{00441B} \color{white} 54\% & \cellcolor[HTML]{F7FBFF} \color{black} 0\% & \cellcolor[HTML]{0B7734} \color{white} 31\% & \cellcolor[HTML]{D6E6F4} \color{black} 2\% & \cellcolor[HTML]{00441B} \color{white} 51\% & \cellcolor[HTML]{F7FBFF} \color{black} 0\% & \cellcolor[HTML]{00441B} \color{white} 62\% & \cellcolor[HTML]{F7FBFF} \color{black} 0\% & \cellcolor[HTML]{005A24} \color{white} 35\% & \cellcolor[HTML]{D6E6F4} \color{black} 2\% & \cellcolor[HTML]{00441B} \color{white} 43\% & \cellcolor[HTML]{F7FBFF} \color{black} 1\% & \cellcolor[HTML]{00441B} \color{white} 54\% & \cellcolor[HTML]{F7FBFF} \color{black} 0\% \\
      \bottomrule
    \end{tabular}
  }%
  \vspace*{-0.1cm}
\end{table*}
\begin{table*}[tb]
  \centering
  \captionof{table}{False-positive (FP) and false-negative (FN) rates for TREC-DL 2022 after multilingual Query Phrase (QP) and Instruction (Instruct) injections with PromptArmor filtering. "BASE" denotes the uninjected baseline and is left blank as there are no attack to mitigate. \vspace*{-0.1cm}}
  \label{tab:fp_fn_total_all_rem}
  \resizebox{0.98\textwidth}{!}{%
    \begin{tabular}{llcccccccccccccccccc}
      \toprule
      \multirow{2}{*}{\textbf{Model}} &          & \multicolumn{2}{c}{\textbf{BASE}}             & \multicolumn{2}{c}{\textbf{AR}}             & \multicolumn{2}{c}{\textbf{ENG}}             & \multicolumn{2}{c}{\textbf{GA}}             & \multicolumn{2}{c}{\textbf{HE}}             & \multicolumn{2}{c}{\textbf{RU}}             & \multicolumn{2}{c}{\textbf{SW}}             & \multicolumn{2}{c}{\textbf{TH}}             & \multicolumn{2}{c}{\textbf{VI}}             \\
      \cmidrule(lr){3-4}\cmidrule(lr){5-6}\cmidrule(lr){7-8}\cmidrule(lr){9-10}\cmidrule(lr){11-12}\cmidrule(lr){13-14}\cmidrule(lr){15-16}\cmidrule(lr){17-18}\cmidrule(lr){19-20}
                                      &          & FP                                          & FN                                         & FP                                          & FN                                         & FP                                          & FN                                         & FP                                          & FN                                         & FP                                          & FN                                         & FP                                          & FN                                         & FP                                          & FN                                         & FP                                          & FN                                         & FP                                          & FN                                         \\
      \midrule

      \multirow{4}{*}{\textbf{GPT-OSS}} & QP       & \cellcolor[HTML]{FFFFFF} \color{black} -- & \cellcolor[HTML]{FFFFFF} \color{black} -- & \cellcolor[HTML]{DBF1D6} \color{black} 20\% & \cellcolor[HTML]{3787C0} \color{white} 5\% & \cellcolor[HTML]{DBF1D6} \color{black} 21\% & \cellcolor[HTML]{6AAED6} \color{black} 4\% & \cellcolor[HTML]{E7F6E3} \color{black} 19\% & \cellcolor[HTML]{08306B} \color{white} 7\% & \cellcolor[HTML]{DBF1D6} \color{black} 20\% & \cellcolor[HTML]{3787C0} \color{white} 5\% & \cellcolor[HTML]{E7F6E3} \color{black} 19\% & \cellcolor[HTML]{3787C0} \color{white} 5\% & \cellcolor[HTML]{E7F6E3} \color{black} 19\% & \cellcolor[HTML]{08306B} \color{white} 7\% & \cellcolor[HTML]{E7F6E3} \color{black} 19\% & \cellcolor[HTML]{3787C0} \color{white} 5\% & \cellcolor[HTML]{E7F6E3} \color{black} 19\% & \cellcolor[HTML]{3787C0} \color{white} 5\% \\
                                       & Instruct & \cellcolor[HTML]{FFFFFF} \color{black} -- & \cellcolor[HTML]{FFFFFF} \color{black} -- & \cellcolor[HTML]{DBF1D6} \color{black} 20\% & \cellcolor[HTML]{105BA4} \color{white} 6\% & \cellcolor[HTML]{73C476} \color{black} 25\% & \cellcolor[HTML]{3787C0} \color{white} 5\% & \cellcolor[HTML]{DBF1D6} \color{black} 20\% & \cellcolor[HTML]{6AAED6} \color{black} 4\% & \cellcolor[HTML]{E7F6E3} \color{black} 19\% & \cellcolor[HTML]{08306B} \color{white} 7\% & \cellcolor[HTML]{E7F6E3} \color{black} 19\% & \cellcolor[HTML]{105BA4} \color{white} 6\% & \cellcolor[HTML]{DBF1D6} \color{black} 20\% & \cellcolor[HTML]{3787C0} \color{white} 5\% & \cellcolor[HTML]{DBF1D6} \color{black} 20\% & \cellcolor[HTML]{3787C0} \color{white} 5\% & \cellcolor[HTML]{DBF1D6} \color{black} 20\% & \cellcolor[HTML]{3787C0} \color{white} 5\% \\
                                       & Distractor & \cellcolor[HTML]{FFFFFF} \color{black} -- & \cellcolor[HTML]{FFFFFF} \color{black} -- & \cellcolor[HTML]{DBF1D6} \color{black} 21\% & \cellcolor[HTML]{105BA4} \color{white} 6\% & \cellcolor[HTML]{88CE87} \color{black} 24\% & \cellcolor[HTML]{3787C0} \color{white} 5\% & \cellcolor[HTML]{EFF9EC} \color{black} 18\% & \cellcolor[HTML]{08306B} \color{white} 8\% & \cellcolor[HTML]{DBF1D6} \color{black} 20\% & \cellcolor[HTML]{105BA4} \color{white} 6\% & \cellcolor[HTML]{DBF1D6} \color{black} 21\% & \cellcolor[HTML]{3787C0} \color{white} 5\% & \cellcolor[HTML]{DBF1D6} \color{black} 20\% & \cellcolor[HTML]{105BA4} \color{white} 6\% & \cellcolor[HTML]{DBF1D6} \color{black} 21\% & \cellcolor[HTML]{105BA4} \color{white} 6\% & \cellcolor[HTML]{BEE5B8} \color{black} 22\% & \cellcolor[HTML]{105BA4} \color{white} 6\% \\
                                       & Variant  & \cellcolor[HTML]{FFFFFF} \color{black} -- & \cellcolor[HTML]{FFFFFF} \color{black} -- & \cellcolor[HTML]{E7F6E3} \color{black} 19\% & \cellcolor[HTML]{105BA4} \color{white} 6\% & \cellcolor[HTML]{A2D99B} \color{black} 23\% & \cellcolor[HTML]{3787C0} \color{white} 5\% & \cellcolor[HTML]{E7F6E3} \color{black} 19\% & \cellcolor[HTML]{105BA4} \color{white} 6\% & \cellcolor[HTML]{DBF1D6} \color{black} 20\% & \cellcolor[HTML]{3787C0} \color{white} 5\% & \cellcolor[HTML]{DBF1D6} \color{black} 20\% & \cellcolor[HTML]{3787C0} \color{white} 5\% & \cellcolor[HTML]{E7F6E3} \color{black} 19\% & \cellcolor[HTML]{3787C0} \color{white} 5\% & \cellcolor[HTML]{DBF1D6} \color{black} 20\% & \cellcolor[HTML]{3787C0} \color{white} 5\% & \cellcolor[HTML]{E7F6E3} \color{black} 19\% & \cellcolor[HTML]{105BA4} \color{white} 6\% \\

      \midrule

      \multirow{4}{*}{\textbf{QWEN}} & QP       & \cellcolor[HTML]{FFFFFF} \color{black} -- & \cellcolor[HTML]{FFFFFF} \color{black} -- & \cellcolor[HTML]{37A055} \color{white} 28\% & \cellcolor[HTML]{ABD0E6} \color{black} 3\% & \cellcolor[HTML]{238B45} \color{white} 29\% & \cellcolor[HTML]{ABD0E6} \color{black} 3\% & \cellcolor[HTML]{238B45} \color{white} 30\% & \cellcolor[HTML]{6AAED6} \color{black} 4\% & \cellcolor[HTML]{238B45} \color{white} 29\% & \cellcolor[HTML]{ABD0E6} \color{black} 3\% & \cellcolor[HTML]{37A055} \color{white} 28\% & \cellcolor[HTML]{ABD0E6} \color{black} 3\% & \cellcolor[HTML]{238B45} \color{white} 29\% & \cellcolor[HTML]{6AAED6} \color{black} 4\% & \cellcolor[HTML]{37A055} \color{white} 28\% & \cellcolor[HTML]{ABD0E6} \color{black} 3\% & \cellcolor[HTML]{37A055} \color{white} 28\% & \cellcolor[HTML]{ABD0E6} \color{black} 3\% \\
                                       & Instruct & \cellcolor[HTML]{FFFFFF} \color{black} -- & \cellcolor[HTML]{FFFFFF} \color{black} -- & \cellcolor[HTML]{238B45} \color{white} 29\% & \cellcolor[HTML]{6AAED6} \color{black} 4\% & \cellcolor[HTML]{238B45} \color{white} 30\% & \cellcolor[HTML]{ABD0E6} \color{black} 3\% & \cellcolor[HTML]{238B45} \color{white} 29\% & \cellcolor[HTML]{ABD0E6} \color{black} 3\% & \cellcolor[HTML]{37A055} \color{white} 28\% & \cellcolor[HTML]{3787C0} \color{white} 5\% & \cellcolor[HTML]{37A055} \color{white} 28\% & \cellcolor[HTML]{3787C0} \color{white} 5\% & \cellcolor[HTML]{238B45} \color{white} 29\% & \cellcolor[HTML]{ABD0E6} \color{black} 3\% & \cellcolor[HTML]{238B45} \color{white} 29\% & \cellcolor[HTML]{ABD0E6} \color{black} 3\% & \cellcolor[HTML]{238B45} \color{white} 29\% & \cellcolor[HTML]{ABD0E6} \color{black} 3\% \\
                                       & Distractor & \cellcolor[HTML]{FFFFFF} \color{black} -- & \cellcolor[HTML]{FFFFFF} \color{black} -- & \cellcolor[HTML]{00682A} \color{white} 33\% & \cellcolor[HTML]{ABD0E6} \color{black} 3\% & \cellcolor[HTML]{005A24} \color{white} 35\% & \cellcolor[HTML]{D6E6F4} \color{black} 2\% & \cellcolor[HTML]{37A055} \color{white} 28\% & \cellcolor[HTML]{6AAED6} \color{black} 4\% & \cellcolor[HTML]{238B45} \color{white} 30\% & \cellcolor[HTML]{6AAED6} \color{black} 4\% & \cellcolor[HTML]{00682A} \color{white} 33\% & \cellcolor[HTML]{ABD0E6} \color{black} 3\% & \cellcolor[HTML]{238B45} \color{white} 30\% & \cellcolor[HTML]{6AAED6} \color{black} 4\% & \cellcolor[HTML]{0B7734} \color{white} 31\% & \cellcolor[HTML]{ABD0E6} \color{black} 3\% & \cellcolor[HTML]{00682A} \color{white} 33\% & \cellcolor[HTML]{ABD0E6} \color{black} 3\% \\
                                       & Variant  & \cellcolor[HTML]{FFFFFF} \color{black} -- & \cellcolor[HTML]{FFFFFF} \color{black} -- & \cellcolor[HTML]{46AE60} \color{black} 27\% & \cellcolor[HTML]{6AAED6} \color{black} 4\% & \cellcolor[HTML]{0B7734} \color{white} 32\% & \cellcolor[HTML]{ABD0E6} \color{black} 3\% & \cellcolor[HTML]{37A055} \color{white} 28\% & \cellcolor[HTML]{6AAED6} \color{black} 4\% & \cellcolor[HTML]{37A055} \color{white} 28\% & \cellcolor[HTML]{6AAED6} \color{black} 4\% & \cellcolor[HTML]{46AE60} \color{black} 27\% & \cellcolor[HTML]{ABD0E6} \color{black} 3\% & \cellcolor[HTML]{46AE60} \color{black} 27\% & \cellcolor[HTML]{6AAED6} \color{black} 4\% & \cellcolor[HTML]{46AE60} \color{black} 27\% & \cellcolor[HTML]{ABD0E6} \color{black} 3\% & \cellcolor[HTML]{46AE60} \color{black} 27\% & \cellcolor[HTML]{6AAED6} \color{black} 4\% \\
      \bottomrule
    \end{tabular}
  }%
  \vspace*{-0.1cm}
\end{table*}

\section{Methodology}
We examine the impact of language-based adversarial injections on relevance
judgments produced by LLM-as-a-judge systems. Our experiments use queries and
passages from the 2022 TREC Deep Learning (DL) tracks. These benchmark
collections are widely regarded as the standard testbeds for evaluating modern
neural retrieval assessment methods, as they provide large-scale, professionally
curated relevance judgments. We evaluated two open-weight LLMs using the
UMBRELA~\cite{upadhyayLargeScaleStudyRelevance2025} and
Criteria-Based~\cite{farziCriteriaBasedLLMRelevance2025} prompting frameworks,
both of which have demonstrated strong effectiveness in previous work.
Adversarial manipulation is introduced through keyword injection
attacks~\cite{alaofiLLMsCanBe2025} and instruction-based injections inspired by
the Kaggle ``Can't Please Them All''
competition.\footnote{\url{https://www.kaggle.com/competitions/llms-you-cant-please-them-all/}}

Injected content is generated in 8 languages covering diverse scripts, writing
directions, and resource levels, with at least one language from each level
defined by \citet{joshiStateFateLinguistic2020}. We evaluate two open-weight
models (GPT-OSS-20B and Qwen3-32B) on TREC-DL 2022 to assess robustness under
multilingual adversarial conditions. Performance is measured using false
positive (FP) and false negative (FN) rates against human judgments, where FP
denotes overestimation and FN underestimation. On non-injected passages,
GPT-OSS-20B achieves 19\% FP and 5\% FN, while Qwen3-32B achieves 23\% FP and
3\% FN.

The most straightforward defense mechanism is rule-based filtering, which
detects and removes keywords commonly associated with prompt injection (e.g.,
``ignore''). However, this approach is increasingly ineffective in modern
settings. Contemporary LLMs exhibit strong cross-lingual capabilities,
enabling adversaries to craft injections in a wide range of languages.
Constructing and maintaining comprehensive keyword filters across all languages
is, therefore, impractical and difficult to scale.

A more advanced approach is to adopt LLM-centric filtering methods, such as
PromptArmor~\cite{shiPromptArmorSimpleEffective2025}, which leverages the model
itself to identify and remove injected content. Due to the same cross-lingual
capabilities, LLMs can effectively identify and eliminate instruction-based
injections regardless of the language used before judging, reducing
false-positive rates to near-baseline levels. However, this defense method is
substantially less effective against content-based injections of the type
described by \citet{alaofiLLMsCanBe2025}. LLMs generally do not identify the
appearance of query phrases and keywords in the passage as a form of prompt
injection. Thus, in most cases, PromptArmor fails to detect and remove such
manipulations. Moreover, manual inspection reveals an important utility
limitation: when no clear injection is present, the model may incorrectly remove
or alter legitimate passage content. This over-sanitization degrades the
integrity of the original text and can introduce FPs in downstream relevance
judgments.

PromptArmor can be further modified to explicitly detect query injection. This
enhanced variant can eliminate nearly all direct query injection attempts.
However, such defenses remain inherently reactive and can be easily circumvented
by adaptive attackers. In particular, query variation, which had been widely
used in the IR community to improve retrieval performance~\cite{benham2019var},
can also be repurposed to strengthen query injection attacks. To examine this,
we generated query variants from the original query using GPT-OSS-20B and
injected them into passages. Preliminary results indicate that a subset of these
injections successfully evade detection, suggesting that LLM-based filters often
rely on surface-level or exact-match cues when identifying injected content.

We further test adaptive attacks. Query variants, generated from the original
query, often evade detection, suggesting reliance on surface cues. We also
introduce distracting content that mimics natural passages while remaining
irrelevant. Following \citet{cuconasuPowerNoiseRedefining2024}, such content is
manually verified to be non-relevant. Our results show that these strategies
bypass existing defenses and can induce false positives, especially in smaller
models.

% \section{Results \& Discussion}
% Our results (Table~\ref{tab:fp_fn_percentage}) show that cross-lingual
% injections consistently inflate FP rates across languages,
% models, and prompting frameworks. We observe a trade-off between FP and FN rates, 
% indicating that while a more strict model can lower the FP rate, it also introduces
% the risk of underestimating the relevance of passages.
% Model behavior also differs: GPT-OSS produces lower FP
% but higher FN rates than Qwen3-32B, suggesting a more conservative strategy.
% This pattern is consistent across languages and injection types.
% Instruction-based injections are effective at inducing FPs but are the easiest
% to mitigate, consistent with prior work and existing defenses
% \cite{perez2022ignore, chenDefensePromptInjection2025,
% shiPromptArmorSimpleEffective2025, li2025survey}.

\section{Results \& Discussion}
Table ~\ref{tab:fp_fn_total_vulgen} shows that cross-lingual injections
consistently increase FP rates across all languages, models, and prompting
frameworks. Compared to non-injected baselines, most injection types either
maintain or further increase FP rates, indicating persistent relevance inflation
under adversarial conditions. We observe a clear FP-FN trade-off: reducing FP
typically increases FN, and vice versa. GPT-OSS exhibits lower FP but higher FN
than Qwen3, suggesting a more conservative strategy, while Qwen3 is more
permissive and prone to overestimation. 

Table~\ref{tab:fp_fn_total_all_rem} shows that PromptArmor reduces attack effectiveness but does not fully eliminate false-positive inflation. Across attacks, instruction-based injections are effective but remain the
easiest to mitigate. In contrast, content-based methods, especially query
variants and distracting content, are more challenging. Query variants
consistently match or exceed baseline FP rates across languages, indicating that
simple transformations are sufficient to preserve attack effectiveness.
Distracting content is the most effective attack, raising FP rates up to 35\%
for Qwen3, while keeping FN low. These effects are stable across languages,
suggesting that multilinguality does not reduce attack effectiveness and may
limit the reliability of language-specific defenses.

\subsection{Statistical Significance}
\begin{table}[httpb]
    \centering
    \caption[ANOVA 2022]{Two-way ANOVA for mean difference of labels with model and language as factors on TREC-DL 2022.}
    \label{tab:anova_language_model_mean_diff_2022}
    \begin{tabular}{l r r r r}
    \toprule
    term & sum squared & df & F & PR(>F) \\
    \midrule
    C(model) & 35.303 & 2 & 19.24 & <0.001 \\
    C(lang) & 232.374 & 10 & 25.33 & <0.001 \\
    C(model):C(lang) & 2661.910 & 20 & 145.1 & <0.001 \\
    Residual & 160313.274 & 174768 &  &  \\
    \bottomrule
    \end{tabular}
\end{table}

To determine whether language has a statistically significant impact on LLM-based relevance judgments, we conducted a two-way ANOVA on the mean difference scores for the QP injection results in TREC-DL 2021 and TREC-DL 2022, treating the LLM model and injected language as independent variables. As shown in Table~\ref{tab:anova_language_model_mean_diff_2022}, both factors have statistically significant effects on LLM-generated relevance labels. Moreover, the interaction effect between model and language is significant in both datasets, indicating that the impact of adversarial language depends on the specific LLM judge being evaluated.

\section{Conclusions}
This study shows that multilingual content-based injections, including query
variants and semantically distracting text represent a key vulnerability in
LLM-as-a-judge systems. Unlike instruction-based attacks, these inputs bypass
standard filtering approaches, such as PromptArmor, consistently inflate
relevance scores across languages.

Future work should examine how such inflated judgments affect downstream
retrieval-augmented generation (RAG) pipelines and whether similar effects
extend to human relevance assessment under exposure to multilingual
content-based injections.

\begin{acks}
  This research is supported by the ARC Centre of Excellence for Automated
  Decision-Making and Society (ADM+S, CE200100005). The experiments reported in
  this paper were undertaken with the assistance of computing resources from
  RACE (RMIT AWS Cloud Supercomputing). We acknowledge the Woi wurrung and Boon
  wurrung language groups of the Kulin Nation as the Traditional Owners of the
  land on which this research was conducted, and pay our respect to Aboriginal
  and Torres Strait Islander peoples and their connection to land and community.
\end{acks}

\bibliographystyle{ACM-Reference-Format}
\balance
\bibliography{references}

\end{document}